\documentclass[aps,pre,twocolumn,groupedaddress]{revtex4}

\usepackage{graphicx} 
\usepackage{dcolumn} 
\usepackage{bm}
\newcommand{\ignore}[1]{} 
\usepackage{epsfig} 
\usepackage{color}
\usepackage{hyperref}
\usepackage{amsmath,amssymb}

\usepackage{comment}
\usepackage[TS1,T1]{fontenc}
\usepackage[latin1]{inputenc}

\begin{document}

\title{The role of voting intention in public opinion polarization}

\author{Federico Vazquez}
\email{fede.vazmin@gmail.com}
\affiliation{Instituto de C\'{a}lculo, FCEN, Universidad de Buenos Aires and CONICET, Buenos Aires, Argentina}
\homepage{https://fedevazmin.wordpress.com}

\author{Nicolas Saintier}
\email{nsaintie@dm.uba.ar}
\affiliation{Departamento de Matem\'atica and IMAS, UBA-CONICET, Facultad de Ciencias Exactas y Naturales, Universidad de Buenos Aires (1428) Pabell\'on I - Ciudad Universitaria - Buenos Aires - Argentina}

\author{Juan Pablo Pinasco}
\email{jpinasco@dm.uba.ar}
\affiliation{Departamento de Matem\'atica and IMAS, UBA-CONICET, Facultad de Ciencias Exactas y Naturales, Universidad de Buenos Aires (1428) Pabell\'on I - Ciudad Universitaria - Buenos Aires - Argentina}

\date{\today}

\begin{abstract}
We introduce and study a simple model for the dynamics of voting intention in a population of agents that have to choose between two candidates.  The level of indecision of a given agent is modeled by its propensity to vote for one of the two alternatives, represented by a variable $p \in [0,1]$.  When an agent $i$ interacts with another agent $j$ with propensity $p_j$, then $i$ either increases its propensity $p_i$ by $h$ with probability $P_{ij}=\omega p_i+(1-\omega)p_j$, or decreases $p_i$ by $h$ with probability $1-P_{ij}$, where $h$ is a fixed step.  We analyze the system by a rate equation approach and contrast the results with Monte Carlo simulations.  We found that the dynamics of propensities depends on the weight $\omega$ that an agent assigns to its own propensity.  When all the weight is assigned to the interacting partner ($\omega=0$), agents' propensities are quickly driven to one of the extreme values $p=0$ or $p=1$, until an extremist absorbing consensus is achieved.  However, for $\omega>0$ the system first reaches a quasi-stationary state of symmetric polarization where the distribution of propensities has the shape of an inverted Gaussian with a minimum at the center $p=1/2$ and two maxima at the extreme values $p=0,1$, until the symmetry is broken and the system is driven to an extremist consensus.  A linear stability analysis shows that the lifetime of the polarized state, estimated by the mean consensus time $\tau$, diverges as $\tau \sim (1-\omega)^{-2} \ln N$ when $\omega$ approaches $1$, where $N$ is the system size.  Finally, a continuous approximation  allows to derive a transport equation whose convection term is compatible with a drift of particles from the center towards the extremes.
\end{abstract}

\maketitle

\section{Introduction}
\label{intro}

Political bi-polarization is a widespread phenomenon that generates divisions in a society, and even clashes and revolts.  Several works try to explain and model the emergence of polarization using different mechanisms, like negative influence between individuals of antagonistic opinion groups or between members of the same group \cite{Baldassarri-2007,Balenzuela-2015,Flache-2011,Salzarulo-2006}, or a confirmation bias, by which individuals tend to search for information that affirms their prior believes and discard arguments that confront their opinions \cite{Anagnostopoulos-2014,Sunstein-2002,Zollo-2015}.  More recently, it has been proposed a new alternative mechanism that gives rise to bi-polarization, which combines homophily with the theory of persuasive arguments \cite{Myers-1982,Mark-2003,Lau-1998}.  The idea is that homophily increases interactions between individuals with the same opinion orientation who then persuade each other with arguments that support their opinion tendency, reinforcing their initial positions and becoming more extreme in their believes (see \cite{Mas-2013,Mas-2013-2,LaRocca-2014,Velasquez-2018}).  The model studied in \cite{Mas-2013} assumes that each agent has a list with a number of pro and con arguments in favor and against a given issue (e.g. marijuana legalization), respectively.  Agents interact by pairs and incorporate in their list of arguments one of its partner's argument chosen at random, while old arguments are dismissed.  Within the context of voting intention, it is natural to think that the number of arguments in favor of a given candidate is proportional to the inclination or propensity to vote for that candidate prior to the elections.

In this article we study the dynamics of propensities in a population of voters that have to decide between two candidates $A$ and $B$.  When two agents meet, the first agent asks its partner about its voting intention, whose answer ($A$ or $B$) depends on its propensity for that candidate.  For simplicity we assume that the second agent answers $A$ with a probability equal to the fraction of its arguments in favor of candidate $A$, that is, its propensity for $A$ (and equivalently for $B$).  If the answer is $A$, then the first agent increases its propensity and thus becomes more prone to voter for $A$.  Otherwise, if the answer is $B$ the propensity of the first agent is decreased and becomes more prone to $B$.  

This model is simple enough to be analytically tractable and is able to induce polarization without relying on the previous mentioned mechanisms, and by implementing a pairwise interaction rule that is independent on the opinion orientation, unlike the models studied in 
\cite{LaRocca-2014,Velasquez-2018}.

The article is organized as follows.  In Section~\ref{model} we introduce the model and its dynamics.  In Section~\ref{rate} we derive the rate equations and obtain their steady state solutions.  We present simulation results in section~\ref{MC}.  An stability analysis of the rate equations is performed in sections~\ref{stability-2} and \ref{stability-S}.  In section~\ref{approximation} we develop a continuous approximation that allows to derive a transport equation.  We conclude in section~\ref{summary} with a short summary and a discussion of the results.

\section{Description of the model}
\label{model}

We consider a population of $N$ interacting voters that have to choose between two candidates, $A$ or $B$.  During the period previous to the election day, agents have some degree of indecision about the two possible alternatives, which is modeled by assuming that each agent has an inclination or propensity $p$ to vote for candidate $A$ that varies between $0$ and $1$, so that when $p \simeq 0$ ($p \simeq 1$) the agent is prone to choose $B$ ($A$).
We assume that the propensity of each agent evolves under the influence of the other agents, in such a way that agents update their propensities after pairwise interactions.  Initially, the propensities of all agents are uniformly distributed in $[0,1]$.  Then, at each time step $\Delta t=1/N$ of the dynamics, two agents $i$ and $j$ are chosen at random to interact.  Agent $j$ tells its partner $i$ that is going to vote for $A$ with a probability equal to its propensity $p_j$.  Then, agent $i$ can either increase or decrease its propensity with probabilities that depend on its own and its partner's propensity: 
\begin{eqnarray}
  p_i(t+1/N) = 
  \begin{cases}
    p_i(t)+h & \mbox{with probability $P_{i,j}$}, \\
    p_i(t)-h & \mbox{with probability $1-P_{i,j}$},
  \end{cases}
  \label{pi}
\end{eqnarray}
where
\begin{equation}
  P_{i,j} = \omega \, p_i(t) + (1-\omega) \, p_j(t).
\end{equation}
The step length $h$ ($0 < h \le 1$) is fixed, while the parameter $\omega$
($0 \le \omega \le 1$) is the weight that gives each agent to its own propensity in an interaction.  The value of $p_i$ is set to $1$ ($0$) when it becomes larger (smaller) than $1$ ($0$), so that propensities are constrained to the interval $[0,1]$.  This time step is repeated ad infinitum.

\section{Rate equations and stationary states}
\label{rate}

At time $t=0$ the distribution of propensities $f(p,0)$ is uniform in $[0,1]$, but after a time of order $1$ agents' propensities adopt discrete values $p=0,h,2h,..,1$.  For the sake of simplicity we can take $h$ such that $S \equiv 1/h$ is an integer number, and thus the propensity adopts discrete values $p=kh$, with $k=0,..,S$.  Then, the propensity distribution can be written as
$f(p,t)=\sum_{k=0}^S n_k(t) \, \delta(p-kh)$, where we define $n_k(t)$ as the
fraction of agents in state $k$ (with propensity $kh$) at time
$t$, where $\sum_{k=0}^S n_k(t)=1$ for all times $t \ge 0$.  Then, the evolution of the system is described by the following rate equations:
\begin{subequations}
  \begin{alignat}{3}
    \label{dndt0}
    \frac{dn_0}{dt} &= \left[ 1-\omega h - (1-\omega) m) \right] n_1 -
    (1-\omega) m \, n_0, \\
    \label{dndtk}
    \frac{dn_k}{dt} &= \left[ \omega h (k-1)+(1-\omega) m
      \right] n_{k-1} - n_k \nonumber \\ &+ \big\{ 1 - \left[ \omega h (k+1)+ (1-\omega) m \right] \big\} n_{k+1} \\ & \mbox{for $1 \le k \le S-1$}, \nonumber \\
    \label{dndtS}
    \frac{dn_S}{dt} &=  \left[ \omega (1-h) + (1-\omega) m
      \right] n_{S-1} - (1-\omega) (1-m) n_S,
  \end{alignat}
  \label{dndt}
\end{subequations}
where $m \equiv \langle p \rangle = h \sum_{k=0}^S k \, n_k$ is the mean value of the propensity.  The first (gain) term in Eq.~(\ref{dndtk}) corresponds to the transition of particles from state $k-1$ to state $k$, which happens with probability $\omega h(k-1)+ (1-\omega) hk'$ when they interact with another particle in state $k'$.  Adding over all $k'$ values leads to $m$.  The second (loss) term represents $k \to k-1$ and $k \to k+1$ transitions, which happen with probability $1$ after interacting with any other particle.  Finally, the third (gain) term is analogous to the first term, where particles switch from state $k+1$ to state $k$ with probability $1-[\omega h(k+1)+(1-\omega)h k']$ when they interact with a $k'$-particle.  

We are interested in the stationary distributions of propensities in the population, which correspond to the fixed points of the system of equations~(\ref{dndt}).  On the one hand, we can first notice that the two consensus states in the extreme propensity values $p=0$ ($n_0^*=1$, $n_k^*=0$ for $k=1,..,S$) and $p=1$ ($n_S^*=1$, $n_k^*=0$ for $k=0,..,S-1$) are fixed points of Eqs.~(\ref{dndt}) that correspond to the two absorbing states of the particle system, where the mean propensities are $m=0$ and $m=1$, respectively.  On the other hand, we show in Appendix~\ref{stat-disc} that for a given mean propensity $m<1$ the system of Eqs.~(\ref{dndt}) has non-trivial fixed points given by
\begin{equation}
  \begin{split}
    n_0^* & =  \frac{1}{1+\sum_{k=1}^S G_k(h,\omega,m)}, \\ 
    n_k^* & =  n_0^* \, G_k(h,\omega,m) \qquad \mbox{for ~ $1 \le k \le S$}, 
    \label{n-stat}
  \end{split}
\end{equation}
with
\begin{eqnarray}
  G_k(h,\omega,m) = \frac{\Pi_{j=0}^{k-1} \left[ (1-\omega)m+j\omega h \right]}
  {\Pi_{j=1}^k \left[ 1-(1-\omega)m -j\omega h \right]}, 
  \label{gkmw} 
\end{eqnarray}
or 
\begin{eqnarray}
  G_k(h,\omega,m)  = \frac{ \Gamma \Big( \frac{(1-\omega)m}{\omega h}+k \Big) \,
	 \Gamma\Big( \frac{1-(1-\omega)m}{\omega h}-k\Big) }
  { \Gamma\Big( \frac{(1-\omega)m}{\omega h}\Big) \,
    \Gamma\Big( \frac{1-(1-\omega)m}{\omega h}\Big)} 
  \label{gkmw-G}
\end{eqnarray}
using the gamma functions for $m>0$ (see Appendix~\ref{stat-disc}), and where the mean propensity must satisfy the relation 
\begin{equation}
  m + \sum_{k=1}^S (m-kh) \, G_k(h,\omega,m)=0.
  \label{m-stat} 
\end{equation}
Notice that for $m=0$ we have from Eq.~(\ref{gkmw}) that $G_k(h,\omega,0)=0$ for $1 \le k \le S$ due to the $j=0$ term in the numerator, and thus we recover the consensus solution $n_0^*=1$.  Note also that Eq.~(\ref{gkmw}) is not valid when $m=1$ because the denominator equals $0$ due to the term $j=S=1/h$. 

We now look for non-consensus stationary states $m \neq 0, 1$.  It is instructive to start analyzing the simplest cases $S=1$ and $S=2$.  For $S=1$ ($h=1$) is $G_1(1,\omega,m) =m/(1-m)$ from Eq.~(\ref{gkmw}), and thus Eq.~(\ref{m-stat}) 
\begin{equation}
  m+(m-1)\, G_1(1,\omega,m)=0
\end{equation}
is satisfied for all values of $m$, that is, each $m$ is a stationary value.  This is because for $S=1$ the propensity model turns to be equivalent to the voter model \cite{Clifford-1973,Holley-1975}, where it is known that the fractions of voters in each state are conserved, i e., $n_0(t)=n_0(0)$ and $n_1(t)=n_1(0)=1-n_0(0)$ for all $t \ge 0$.  Then, the stationary value of $m$ is equal to its initial value $m(0)=n_1(0)$ given by the initial propensity distribution.  For $S=2$ ($h=1/2$), Eq.~(\ref{m-stat}) for $m$ is 
\begin{eqnarray}
  m+(m-1/2)\, G_1(h,\omega,m) + (m-1) \, G_2(h,\omega,m)=0.
  \label{m2-stat}
\end{eqnarray}
Replacing in Eq.~(\ref{m2-stat}) the expressions for $G_1$ and $G_2$
\begin{subequations}
  \begin{alignat}{2}
    G_1(h,\omega,m) &= \frac{(1-\omega)m}{1-\omega/2-(1-\omega)m} ~~~\mbox{and} \\ G_2(h,\omega,m) &= \frac{m \left[(1-\omega)m+\omega/2 \right]}{(1-m)\left[1-\omega/2-(1-\omega)m \right]},
  \end{alignat}
  \label{G1-G2}
\end{subequations}
we arrive, after doing some algebra, to the simple relation
\begin{eqnarray*}
  (1-\omega)m(1-m)(1-2m)= 0. 
\end{eqnarray*}
Therefore, besides the extreme consensus states $m^*=0$ and $m^*=1$, we obtain the new stationary solution $m^*=1/2$.  For general $S$, the possible stationary values of $m$ are given by the solutions of Eq.~(\ref{m-stat}), which are the roots of a polynomial of order $S+1$.  We were unable to find the roots of that polynomial analytically for any $S \ge 3$.  However, we have verified numerically for many different values of $\omega$ and $h$ that the only real roots are $m^*=0$, $m^*=1$ and $m^*=1/2$, as in the case $S=2$.

\begin{figure}[t]
  \includegraphics[width=\columnwidth]{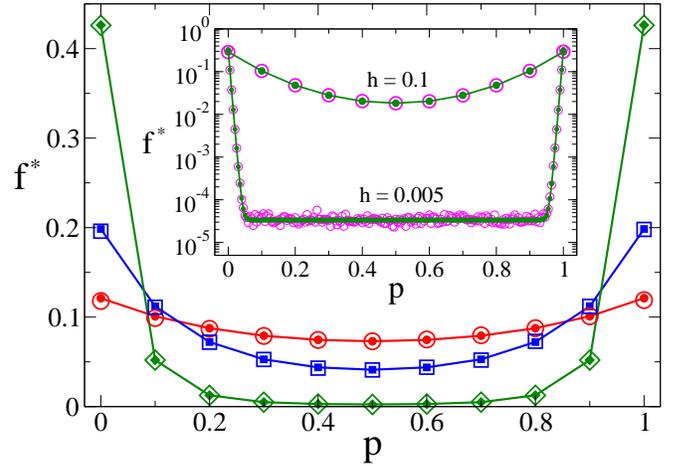}
  \caption{Main: distribution of propensities at the stationary polarized state for step $h=0.1$ and weights $\omega=0.1$ (circles), $\omega=0.3$ (squares) and $\omega=0.8$ (diamonds).  Filled symbols joined with lines are the stationary solution Eqs.~(\ref{nk-series1}), while open symbols correspond to MC simulation results.  Inset: upper (lower) curve corresponds to $\omega=0.5$ and $h=0.1$ ($h=0.005$), respectively.}
  \label{f-p-w}
\end{figure}

\begin{figure}[t]
  \includegraphics[width=\columnwidth]{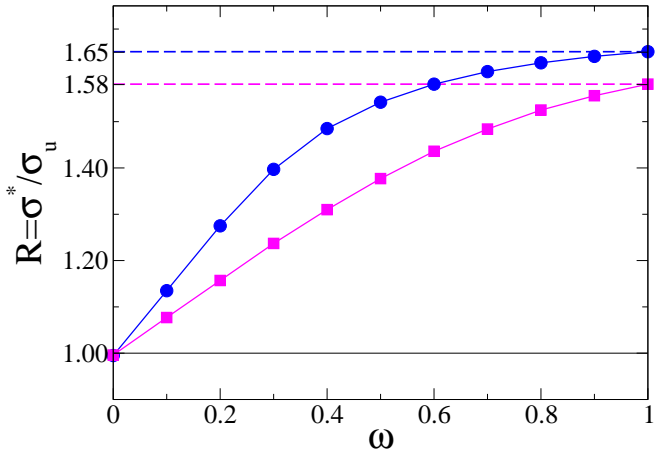}
  \caption{Polarization level $R$ versus weight $\omega$ for $h=0.05$ (circles) and $h=0.1$ (squares).  The horizontal dashed lines denote the saturation values at $\omega=1$.}
  \label{sigma-w}
\end{figure}

We now analyze the non-trivial solution $m^*=1/2$ that corresponds, as we shall see bellow, to a stationary distribution of propensities $f^*(p)$ that is symmetric and polarized around $p=1/2$.  For $m=1/2$, Eqs.~(\ref{n-stat}) and (\ref{gkmw-G}) become
\begin{subequations} 
  \begin{alignat}{3}
    n_{P,0}^* &= \frac{1}{1+\sum_{k=1}^S G_k(h,\omega,1/2)},  \\
    n_{P,k}^* &= n_{P,0}^* \, G_k(h,\omega,1/2) \qquad \mbox{$1 \le k \le S$},  \\
    G_k(h,\omega,1/2) &= \frac{\Gamma\left(\frac{1-\omega}{2\omega h}+k \right) \, \Gamma \left(\frac{1+\omega}{2\omega h}-k \right)}
    {\Gamma \left(\frac{1-\omega}{2\omega h} \right) \, \Gamma \left(\frac{1+\omega}{2\omega h} \right)},
  \end{alignat}
  \label{nk-series1}
\end{subequations}
where the subindex $P$ in Eqs.~(\ref{nk-series1}) stands for polarization.  This series solution Eqs.~(\ref{nk-series1}) is plotted in Fig.~\ref{f-p-w} for $S=10$ ($h=0.1$) and different values of $\omega$ (filled symbols), while the inset shows the behavior for two values of $h$ (filled circles).  We observe that, for each $\omega$, the shape of $f^*(p)$ is symmetric around $p=1/2$ and peaked at the opposite extreme values $p=0$ and $p=1$.  This describes a situation in which propensities in the population are polarized, where most individuals adopt opposite and extreme propensity values.  In the main plot we see that the system becomes more polarized as $\omega$ increases, while we observe in the inset that the polarization is more pronounced as $h$ decreases.  To quantify the level of polarization we computed the ratio $R(h,\omega) \simeq \sigma^*(h,\omega)/\sigma_u$ between the standard deviation  $\sigma^*(h,\omega)=\sqrt{\langle p^2 \rangle - \langle p \rangle^2} = \sqrt{\left( h^2 \sum_{k=0}^S k^2 \, n_k^* \right) - 1/4}$ of the propensity distribution $f^*(p)$ for given values of $h$ and $\omega$, and the corresponding value $\sigma_u=\sqrt{1/12+h/6}$ of the uniform distribution $f_u(p)=\frac{1}{S+1} \sum_{k=0}^S \delta(p-kh)$.  As the width of $f^*(p)$ increases respect to the uniform distribution when the system is polarized, we expect $R>1$ and proportional to the magnitude of the polarization.  Results are shown in Fig.~\ref{sigma-w}.  We see that, for a given $h$, $R$ increases with $\omega$ from the value $1$ for $\omega=0$ to the value $1/(2 \sigma_u)$ corresponding to the double-peak distribution $f(p)=[\delta(p)+\delta(p-1)]/2$ obtained when $\omega=1$.  It is quite remarkable that the system becomes polarized ($R>1$) for any $\omega>0$.  In other words, there is no polarization when agents assign zero weight to it own propensity, but a tiny amount of weight is enough to polarized the population.  In summary, for any $h$ and $\omega$, the only stationary states predicted by the rate equations  (\ref{dndt}) are the consensus absorbing states $m^*=0$ and $m^*=1$, and the symmetric polarized state $m^*=1/2$ in which most agents hold extreme propensities.  This polarization phenomenon appears when $\omega>0$ and is magnified as $\omega$ increases. 

\section{Monte Carlo simulations}
\label{MC}

\begin{figure}[t]
  \includegraphics[width=\columnwidth]{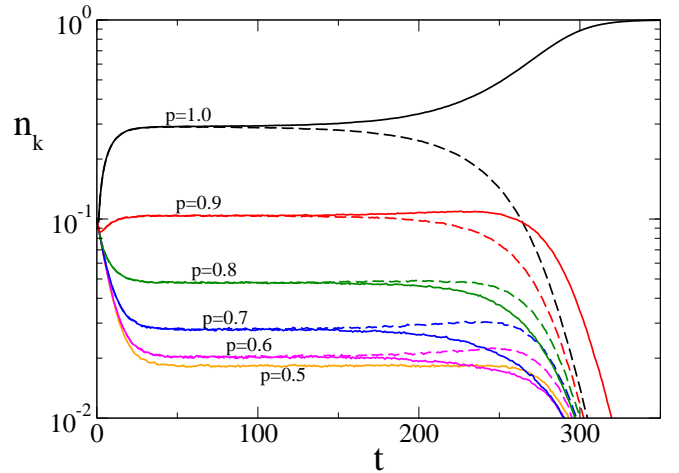}
  \caption{Monte Carlo results for the time evolution of the fraction of agents $n_k$ with propensities $p=k h$ ($k=1,..,10$), with $h=0.1$ and $\omega=0.5$, in a population of $N=10^6$ agents.  Solid (dashed) curves correspond to $p=1.0$ ($0.0$), $p=0.9$ ($0.1$),  $p=0.8$ ($0.2$), $p=0.7$ ($0.3$), $p=0.6$ ($0.4$) and $p=0.5$ (from top to bottom).}
  \label{nk-t}
\end{figure}

To compare the previous analytical results with that obtained from Monte Carlo (MC)  simulations of the model we studied the time evolution of the fractions $n_k$ in single realizations of the dynamics starting from a uniform distribution, as we show in Fig.~\ref{nk-t} for a population of $N=10^6$ agents, $S=10$ ($h=0.1$) and $\omega=0.5$.  We can see that after a short initial transient the fractions $n_k$ reach a nearly constant value (plateau) that depends on $k$.  However, this state in not stable and eventually all $n_k$ finally decay to zero except for $n_{10}$ ($p=1$) that increases and reaches $1$, corresponding to a consensus in $p=1$.  The height of these plateaus are plotted by open symbols in the main panel of Fig.~\ref{f-p-w} for $h=0.1$ and different values of $\omega$, and in the inset of the same figure for $\omega=0.5$ and two values of $h$.  We observe a very good agreement with the analytic stationary values Eqs.~(\ref{nk-series1}) obtained from the rate Eqs.~(\ref{dndt}) (filled symbols), which describe an infinite large system where finite-size fluctuations are neglected.  We have checked that, indeed, the length of the plateaus increase with $N$ and thus they become infinitely large in the thermodynamic limit, corresponding to the stationary states predicted by the theory.  Therefore, the polarized state seems to be unstable, and an extremist consensus is eventually achieved in a finite system.

\begin{figure}[t]
  \includegraphics[width=\columnwidth]{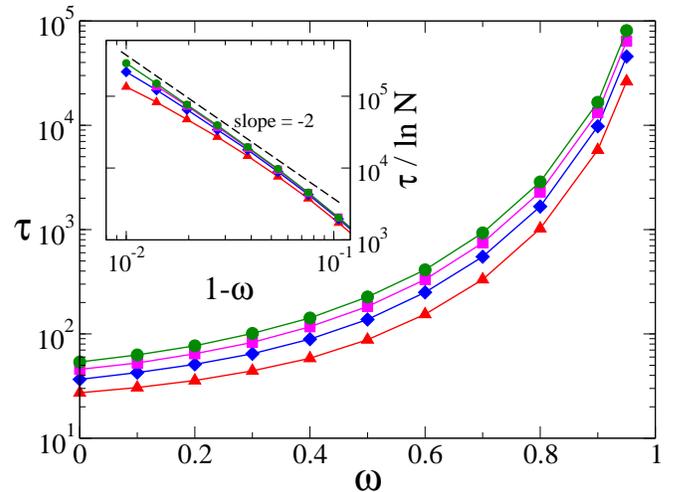}
  \caption{Mean consensus time $\tau$ vs $\omega$ for $h=0.1$ and system sizes $N=20$ (triangles), $N=80$ (diamonds), $N=320$ (squares) and $N=1280$ (circles).  Inset: the data collapse shows the approximate scaling $\tau \sim (1-\omega)^{-2} \ln N$ for $\omega \lesssim 1$.  The dashed line has slope $-2$.}
  \label{tau-w}
\end{figure}

To study the lifetime of the quasiestationary polarized state in finite systems we computed the time to reach consensus.  In Fig.~\ref{tau-w} we show MC results of the mean consensus time $\tau$ vs $\omega$ for various system sizes $N$.  We see that $\tau$ increases with $\omega$ and seems to diverge when $\omega$ approaches $1$ as a power law $\tau \sim (1-\omega)^{-\alpha}$, with $\alpha \simeq 2$ (see inset).  This means that polarization not only gets stronger as $\omega$ increases, but also lasts for longer times.  The collapse of the data in the inset also shows that $\tau$ increases very slowly with $N$, as $\ln N$.  In the next section we perform a linear stability analysis for the $S=2$ case that allows to obtain the exponent $\alpha$ and the logarithmic scaling with $N$.

\section{Stability analysis for the $S=2$ case}
\label{stability-2}

An insight into the results shown in the last section can be obtained by studying the simplest non-trivial case $S=2$, for which the rate equations~(\ref{dndt}) are
\begin{subequations}
  \begin{alignat}{3}
    \label{dndt0-2}
    \frac{dn_0}{dt} &= \left[ 1-\omega/2 - (1-\omega) m) \right] n_1 -
    (1-\omega) m \, n_0, \\
    \label{dndt1-2}
    \frac{dn_1}{dt} &= (1-\omega) m \, n_0 - n_1 + (1-\omega)(1-m) n_2, \\
    \label{dndt2-2}
    \frac{dn_2}{dt} &=  \left[ \omega/2 + (1-\omega) m
      \right] n_1 - (1-\omega) (1-m) n_2.
  \end{alignat}
  \label{dndt-2}
\end{subequations}
It proves convenient to work with the closed system of equations for $n_0$ and $n_2$
\begin{subequations}
  \begin{alignat}{2}
    \label{dndt0-3}
    \frac{dn_0}{dt'} &= \left[ 1 + \epsilon (n_0-n_2) \right] (1-n_0-n_2) -
     \epsilon (1-n_0+n_2) n_0, \\
    \label{dndt2-3}
    \frac{dn_2}{dt'} &= \left[ 1 - \epsilon (n_0-n_2) \right] (1-n_0-n_2) -
    \epsilon (1+n_0-n_2) n_2,
  \end{alignat}
  \label{dndt-3}
\end{subequations}
obtained from Eqs.~(\ref{dndt-2}) by using the identities $n_1=1-n_0-n_2$ and $2m=n_1+2n_2=1-n_0+n_2$ to express $n_1$ and $m$ in terms of $n_0$ and $n_2$, and defining the parameter $\epsilon \equiv 1-\omega$ and the rescaled time $t'=t/2$.  As we proved in section~\ref{rate} for $S=2$, the rate equations have three fixed points.  The two trivial fixed points that represent consensus states in $p=0$ and $p=1$ are $\vec{n}_0^*=(1,0,0)$ and $\vec{n}_2^*=(0,0,1)$, respectively.  The non-trivial fixed point $\vec{n}_{P}^*=(n_{P,0}^*,n_{P,1}^*,n_{P,2}^*)$ corresponding to polarization is
\begin{equation}
  \vec{n}_{P}^*= \left(
  \frac{1}{3-\omega},\frac{1-\omega}{3-\omega},\frac{1}{3-\omega} \right),
  \label{n12}
\end{equation}
which is calculated from Eqs.~(\ref{n-stat}) using the expressions  
$n_{P,0}^*=1/(1+G_1+G_2)$, $n_{P,1}^*=G_1 \, n_{P,0}^*$ and $n_{P,2}^*=G_2 \, n_{P,0}^*$, with $G_1(1/2,\omega,1/2)=1-\omega$ and $G_2(1/2,\omega,1/2)=1$ obtained from Eqs.~(\ref{G1-G2}) for $S=2$ ($h=1/2$) by plugging $m=1/2$.

To investigate how the system approaches consensus we start by performing a linear stability analysis of the trivial fixed point $\vec{n}_0^*=(1,0,0)$ which, by symmetry, is analogous to the analysis of $\vec{n}_2^*$.  We consider small independent perturbations $0 < x_0, x_2 \ll 1$ of the fixed point components and write $n_0=1-x_0$ and $n_2=x_2$.  Plugging these expressions for $n_0$ and $n_2$ into Eqs.~(\ref{dndt-3}) we obtain, to first order in $x_0$ and $x_2$ (neglecting terms of order $2$), the following system of linear equations written in matrix representation:
\begin{eqnarray*}
  \frac{d {\mathbf x}}{dt'} = {\bf A \, x},
\end{eqnarray*}
where  
\begin{eqnarray*}
  {\bf A} \equiv \left( {\begin{array}{cc}
   -1 & 1+2\epsilon \\
   1-\epsilon & -1-\epsilon \\
  \end{array} } \right),
\end{eqnarray*}
and ${\bf x} \equiv (x_0,x_2)$.  The eigenvalues of the matrix ${\bf A}$ \begin{eqnarray}
  \lambda_{\pm} = \frac{-2-\epsilon \pm \sqrt{(2+\epsilon)^2 - 8 \epsilon^2}}{2}
  \label{lambdas}
\end{eqnarray}
are both negative, and thus the fixed point $\vec{n}_0^*$ is stable under a small perturbation in any direction.  To study the behavior of the system for $\omega \lesssim 1$ we expand Eqs.~(\ref{lambdas}) to leading order in $0 < \epsilon \ll 1$.  This gives
\begin{eqnarray}
  \lambda_{+} = - \epsilon^2 + \mathcal O(\epsilon^3), ~~   
  \lambda_{-} = - 2 + \mathcal O(\epsilon) 
\end{eqnarray}
and, therefore, we have 
\begin{eqnarray*}
  x_0(t') &\simeq& a \, e^{-\epsilon^2 t'} + b \, e^{-2t'} ~~ \mbox{and} ~~ \\
  x_2(t') &\simeq& c \, e^{-\epsilon^2 t'} + d \, e^{-2t'},  
\end{eqnarray*}
where $a,b,c$ and $d$ are constants given by the initial condition.  At long times, only the term corresponding to the lowest eigenvalue $\lambda_+$ survives, and thus the time evolution of $n_0$ and $n_2$ after a perturbation from $\vec{n}_0^*$ are approximately given by
\begin{eqnarray}
  n_0(t) \simeq 1 - a \, e^{-\epsilon^2 t/2} ~~ \mbox{and} ~~ n_2(t) \simeq c \, e^{- \epsilon t/2}.
  \label{n0-n2}
\end{eqnarray}
The mean time to reach consensus in a population of $N$ agents can be estimated as the time for which the fraction of agents with propensity $p=0$ becomes larger than $1-1/N$ (less than one agent with propensity $p > 0$).  Then, from Eq.~(\ref{n0-n2}) we obtain that at consensus time $\tau$ is $n_0(\tau)=1-a \, e^{-\epsilon^2 \tau/2}=1-1/N$, from where we arrive to the approximate expression
for the mean consensus time
\begin{equation}
  \tau \simeq \frac{2 \ln(a N)}{(1-\omega)^2}, 
  \label{tau}
\end{equation}
after replacing back $\epsilon$ by $1-\omega$.  The $(1-\omega)^{-2}$ divergence of $\tau$ as $\omega \to 1$ predicted by Eq.~(\ref{tau}) is in good agreement with the exponent $\alpha \simeq 2$ found from MC simulations (see inset of Fig.~\ref{tau-w}).  Equation~(\ref{tau}) also agrees with the logarithmic increase of $\tau$ with $N$ observed in the inset of Fig.~\ref{tau-w}.

\begin{figure}[t]
  \includegraphics[width=6.5cm]{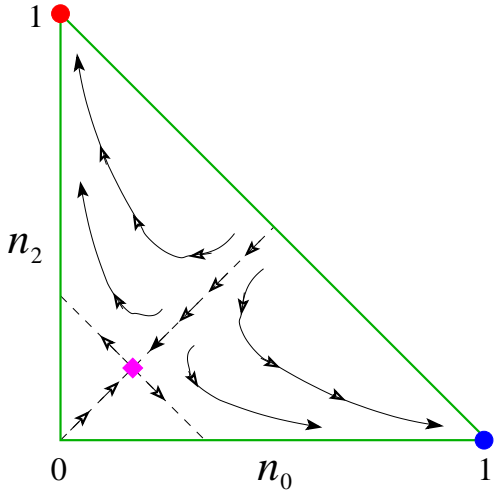}
  \caption{Schematic flow diagram in the $n_0-n_2$ plane for the $3$--propensity system ($S=2$).  The two stable fixed points denoted by circles correspond to the absorbing consensus states in an extreme propensity, while the saddle point denoted by a diamond represents the steady-state of polarization.  The lines with arrows show the flow direction of the system inside the composition triangle $0 \le n_0+n_2 \le 1$.}
  \label{diagram}
\end{figure}

We now study the stability of the polarized state.  For that, we linearize the system of Eqs.~(\ref{dndt-3}) around the fixed point $\vec{n}_{P}^*$ by rewriting $n_0$ and $n_2$ as $n_0=n_{P,0}^*+x_0$ and $n_2=n_{P,2}^*+x_2$, with $0 < |x_0|,|x_2| \ll 1$ and $n_{P,0}^*=n_{P,2}^*=1/(2+\epsilon)$ from Eq.~(\ref{n12}).  Expanding the resulting equations to first order in $x_0$ and $x_2$ we arrive to the following linear system:
\begin{eqnarray*}
  \frac{d {\mathbf x}}{dt''} = {\bf B \, x},
\end{eqnarray*}
with  
\begin{eqnarray*}
  {\bf B} \equiv \left( {\begin{array}{cc}
      -2(1+\epsilon) & -2(1+\epsilon) - \epsilon^2 \\
      -2(1+\epsilon) - \epsilon^2 & -2(1+\epsilon)      
  \end{array} } \right),
\end{eqnarray*}
and $t''=t/(4+2\epsilon)$.  The eigenvalues of ${\bf B}$ are
\begin{eqnarray}
  \lambda_{+} = \epsilon^2 ~~ \mbox{and} ~~
  \lambda_{-} = - 4(1+\epsilon)-\epsilon^2,
  \label{lambdas-1}
\end{eqnarray}
while the associated eigenvectors are
\begin{eqnarray*}
  v_+ = (1,-1) ~~ \mbox{and} ~~ v_-= (1,1).
\end{eqnarray*}
In Fig.~\ref{diagram} we show the flow diagram that summarizes the stability analysis of the $S=2$ case.  The fixed points $\vec{n}_0^*$ and $\vec{n}_2^*$ (circles) are stable in any direction, while $\vec{n}_{P}^*$ (diamond) is a saddle point that is stable only along the $v_-$ direction ($\lambda_- <0$) and unstable along any other direction.  This means that starting from a state that is symmetric around $p=1/2$ [$n_0(0)=n_2(0)$] the system evolves along the line $n_0(t)=n_2(t)$ towards the fixed point $\vec{n}_{P}^*$.  However, any perturbation from the fixed point $\vec{n}_{P}^*$ that is not symmetric around the center propensity $p=1/2$ leads the system to one of the absorbing consensus configurations $p=0$ or $p=1$ for all agents.  This explains the MC simulation results shown in section~\ref{MC}, where the fractions $n_k$ show an initial fast approach from a uniform (symmetric) distribution $n_k(0) \simeq 1/(S+1)$ (with $k=0,..,S$) to the polarized stationary state $\vec{n}_{P}^*$ (see Fig.~\ref{nk-t}) but, eventually, finite-size fluctuations allow the system to escape from this unstable state and reach consensus.

\section{Stability analysis for the $S \ge 3$ case}
\label{stability-S}

In this section we analyze the stability of the fixed points of the rate equations~(\ref{dndt}) for general $S$, namely, the two consensus points $\vec{n}_0^*=(1,0,..,0)$ and $\vec{n}_S^*=(0,0,..,1)$, and the symmetric point $\vec{n}_{P}^*$ given by Eqs.~(\ref{nk-series1}).  As we showed in the last section for the $3$-propensity system ($S=2$), any symmetric distribution evolves towards the fixed point $\vec{n}_P^*$, and one can guess that this behavior also holds for any $S \ge 2$.  More interestingly, we found that for $S\ge 3$ there are non-trivial distributions that are not symmetric around $p=1/2$ which also evolve towards $\vec{n}_P^*$, as we shall see bellow.  We start by reducing the number of independent variables to $S$ using the relation $n_S=1-\sum_{k=0}^{S-1} n_k$ and expressing the mean propensity as
\begin{eqnarray*}
 m = \sum_{k=1}^{S-1} k h \, n_k + n_S= 1 - \sum_{k=0}^{S-1} (kh-1) n_k, \\ 
\end{eqnarray*}
and so we can rewrite Eqs.~(\ref{dndt}) as $\frac{d}{dt}n_k = F_k(n_0,..,n_{S-1})$, where the functions $F_k$ correspond to the right-hand-side of the rate equations.  We can then differentiate $F_k$ around the fixed points to obtain a linear system of equations defined by a linearized matrix ${\bf M}$.  For practical reasons, we used \emph{Mathematica} to calculate the matrix ${\bf M}$, its eigenvalues and eigenvectors.

\begin{figure}[t]
\includegraphics[width=\columnwidth]{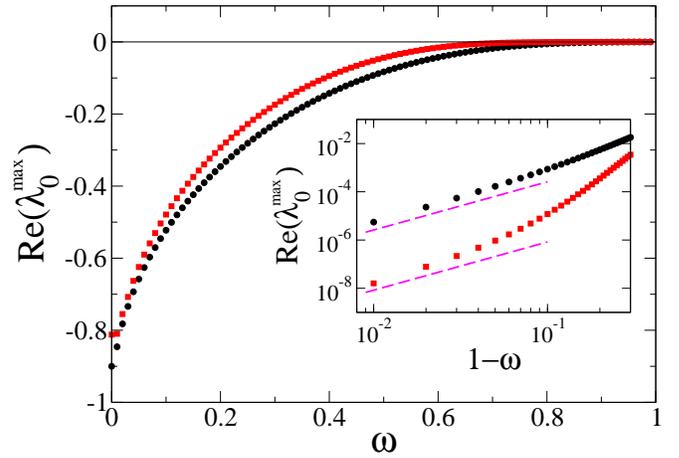}
\caption{Maximum real part of the eigenvalues of matrix ${\bf M_P}$ vs $\omega$ for $h=0.1$ (circles) and $h=0.05$ (squares).  The inset shows how $Re(\lambda)$ approaches $0$ from bellow as $\omega$ goes to $1$.}
\label{eigen-w}
\end{figure}

Let us first analyze the stability of $\vec{n}_0^*$ (the same results hold from the analysis of $\vec{n}_S^*$).  In Fig.~\ref{eigen-w} we plot the maximum of the real part of the eigenvalues $\lambda_0^{\mbox{\tiny max}}$ of ${\bf M_0}$,  calculated numerically, as a function of $\omega$ for two values of $h$.  We can see that $\lambda_0^{\mbox{\tiny max}} < 0$ for all $0 \le w \le 1$.  Therefore, all eigenvalues of ${\bf M_0}$ have a negative real part and thus the consensus fixed point $\vec{n}_0^*$ is locally asymptotically stable.  This result generalizes the stability of the two consensus states found for $S=2$ (section~\ref{stability-2}) to all values $S \ge 2$.  

\begin{figure}[t]	
  \includegraphics[width=\columnwidth]{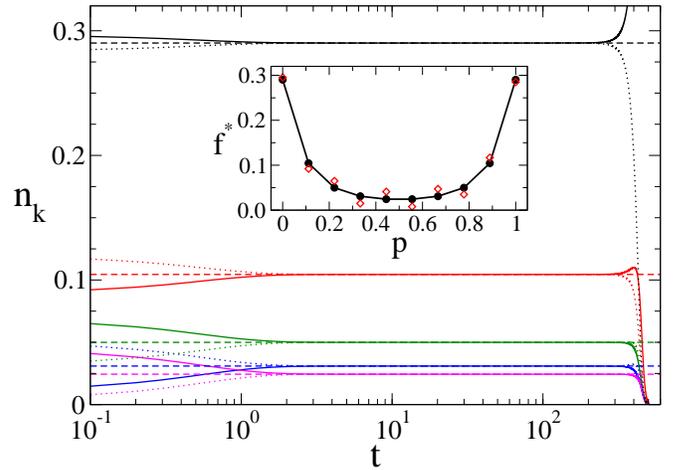}
  \caption{Time evolution of the propensity fractions $n_k$ for a system with   $h=1/9$ ($S=9$) and $\omega=0.5$.  Solid curves correspond to $p=0, 1/9, 2/9, 3/9$ and $4/9$, while dotted curves are for $p=1, 8/9, 7/9, 6/9$ and $5/9$ (from top to bottom).  Inset: the initial values $n_k(0)$ (empty diamonds) correspond to a small perturbation of the polarized stationary distribution given by the components of the fixed point $\vec{n}_P$.}    
  \label{nk-t-pert}
\end{figure}

We now repeat the analysis above for the symmetric fixed point $\vec{n}_{P}^*$.  We observed numerically that, for various values of $S$ and $\omega$, the matrix ${\bf M_P}$ has only one positive eigenvalue and $S-1$ negative eigenvalues.  As we know from standard dynamical system theory, the eigenvalues of ${\bf M_P}$ with negative real part generate the tangent plane $\mathcal T$ to the stable manifold of $\vec{n}_P^*$, which therefore has dimension $S-1$. Besides, the space of propensity distribution $(n_0,..,n_S)$ with mean $m=1/2$ is an affine plane of codimension $1$ which contains $\vec{n}_P^*$, whose intersection with $\mathcal T$ is a manifold of positive dimension $S-1$.  When the system starts from a point of this manifold it follows a trajectory that converges to $\vec{n}_P^*$, that is, the points on $\mathcal T$ represent propensity distributions with mean $m=1/2$ that evolve towards the polarized state.  To illustrate with an example, one of the eigenvectors of ${\bf M_P}$ that we found numerically for case $S=9$ and $\omega=1/2$ is 
\begin{eqnarray*}
\vec{V} \simeq (0.31551,-0.74694,0.90799,-0.97402, 1.00000, \\ -1.00000,0.97402,-0.90799, 0.74694,-0.31551),
\end{eqnarray*}  
whose associated eigenvalue is $\lambda = -1.878148$.  We can now consider
the point $\vec{n}(0) = \vec{n}_P^* + 0.02 \, \vec{V}$ on the plane $\mathcal T$
as a initial state of the system, which is obtained by slightly perturbing the fixed point $\vec{n}_P^*$ in the direction of $\vec{V}$.  The time evolution of the components $n_k$ of $\vec{n}$ are plotted in the main panel of Fig.~\ref{nk-t-pert}, while the inset shows the initial perturbed state (empty diamonds) as compared to $\vec{n}_P^*$ (filled circles).  We can see that the 
$\vec{n}(0)$ is not symmetric with respect to $p=0.5$.  This asymmetry is the result of the components of $\vec{V}$, which exhibit an anti-symmetry that is necessary to preserve the normalization condition $\sum_{k=0}^S n_k(t)=1$ for all $t \ge 1$.  We observe that the fractions $n_k$ (solid and dotted lines) quickly converge to the corresponding values of the components of $\vec{n}_P$ denoted by horizontal dashed lines. However, if we zoom in we can see that $\vec{n}$ gets extremely close to $\vec{n}_P$ but not exactly to $\vec{n}_P$.  This very tiny difference is a consequence of the fact that the initial state $\vec{n}(0)$ belongs to the tangent plane $\mathcal T$ to the stable manifold of $\vec{n}_P$, but a priory not to the stable manifold itself.  Therefore, the systems spends some time very near to $\vec{n}_P$ before eventually going away and converging to the consensus state $p=0$.

\section{Continuous approximation}
\label{approximation}

In order to analyze the polarized state in more detail it proves useful to consider the system of rate Eqs.~(\ref{dndt}) in the limiting case of a very small step $h \ll 1$.  This allows to derive continuous in $p$ partial differential equations that describe the long-time behavior of the system, as we shall see in this section.

As we explained in section~\ref{rate}, after a short initial transient all agents take discrete propensities in the set $p=kh$, with $k=0,..,S$, and thus the propensity distribution can be written as 
\begin{equation}
  f(p,t) = \sum_{k=0}^S n_k(t) \, \delta(p-kh), 
  \label{Measureh} 
\end{equation}
where $\delta(p-kh)$ is the Dirac delta function at $kh$.  Notice that the consensus states correspond to $f(p)=\delta(p)$ and $f(p)=\delta(p-1)$.  We consider a generic function $\phi(p)$ of the propensity, whose mean value over the population is defined as
\begin{equation}
  \langle \phi \rangle_f(t) \equiv \int_0^1 \phi(p) f(p,t) dp =
  \sum_{k=0}^S n_k(t) \, \phi(kh). 
\end{equation}
This is a macroscopic scalar variable of the particle system, like the mean propensity $m(t)=\langle p \rangle(t)$ and its variance when we take $\phi(p)=p$ and $\phi(p)=(p-\langle p \rangle)^2$, respectively.  In Appendix~\ref{continuum-eq} we show that the time evolution of $\langle \phi \rangle_f$ is described by the following equation:
\begin{eqnarray}
  \label{dphidt}
  \frac{1}{h}\frac{d}{dt} \langle \phi \rangle_f &=& 
  \Big \langle v(p,t) \, \phi'(p) + \frac{h}{2}\phi''(p) \Big \rangle_f \nonumber \\
  &+& \left[ 1-B_0(t) \right] \, f(0,t) \left[ \phi'(0) - \frac{h}{2} \phi''(0) \right] \\
  &-& \left[ 1-B_1(t) \right] \, f(1,t) \left[ \phi'(1) + \frac{h}{2} \phi''(1) \right] + \mathcal O(h^2), \nonumber 
\end{eqnarray}
where 
\begin{eqnarray*}
  v(p,t) &\equiv& 2m(t)-1 + 2 \omega \left[ p-m(t) \right], \\
  B_0(t) &\equiv& (1-\omega) m(t) ~~ \mbox{and} \\
  B_1(t) &\equiv& (1-\omega) \left[ 1-m(t) \right].
\end{eqnarray*}
There are no $\mathcal O(h^2)$ terms when $\phi$ is linear in $p$.  As we can see in the derivation of Appendix~\ref{continuum-eq}, the first term in the rhs of Eq.~(\ref{dphidt}) comes from the rate equations for $n_k(t)$ ($0<k<S$) that describe the evolution of the propensity distribution in $(0,1)$, while the second and third terms come from the dynamics near the boundary points at $p=0$ and $p=1$, respectively, and describe the balance between the particles entering and leaving the boundary.  The coefficient $v(p,t)$ is related to the drift of the particles towards the ends of the interval $[0,1]$, while $B_0(t)$  and $B_1(t)$ are boundary coefficients.

Taking $\phi(p)=1$ in Eq.~(\ref{dphidt}) leads to the conservation of the total mass $\int_0^1 f(p,t) dp=1$, as expected.  Besides, for $\phi(p)=p$ we obtain the following equation for the evolution of the mean propensity:
\begin{eqnarray}
  \frac{1}{h}\frac{d}{dt}m(t) &=& 2m(t)-1 + f(0,t) \left[ 1-(1-\omega) m(t) \right] \nonumber \\ &-& f(1,t) \left[ \omega + (1-\omega)m(t) \right].
  \label{dmdt} 
\end{eqnarray}
We can check from Eq.~(\ref{dmdt}) that if the population is initially in a consensus state, i.e. (i) $m(0)=0$ or (ii) $m(0)=1$, then (i) $m(t)=0$ or (ii) $m(t)=1$ for any $t \ge 0$, meaning that the population remains in the consensus state as expected from the fixed point solutions $m^*=0,1$.  We can also see in Eq.~(\ref{dmdt}) that term $2m-1$ describes a drift towards $m=0$ ($m=1$) when $m<1/2$ ($m>1/2$) caused by the instability of the fixed point $m^*=1/2$.  Therefore, starting from a nearly uniform distribution with $m(0)$ slightly larger than $1/2$ as in the MC simulations, agents' propensities are slowly dragged to $p=1$.    
Besides, we see that $m(0)=m^*=1/2$ is a stationary value if $f(0,0)=f(1,0)$, in agreement with the fact that any symmetric distribution $f(1/2-p)=f(1/2+p)$ evolves towards the polarized fixed point $\vec{n}_P^*$, as shown in sections~\ref{stability-2} and \ref{stability-S}.

To better explore the dynamics, we can derive an approximate equation for the time evolution of the propensity distribution $f(p,t)$.  For that, we can rewrite Eq.~(\ref{dphidt}) neglecting order $h$ terms as
\begin{equation}
  \frac{1}{h} \frac{d}{dt} \langle\phi\rangle_f = \int_0^1 [v(p,t)+u(p,t)]f(p,t)\phi'(p)\,dp,\
\end{equation}
where we have introduced the field 
\begin{eqnarray*}
  && u(p,t) = [1-B_0(t)] \delta(p) - [1-B_1(t)]\delta (p-1) \\
  &&= [1-(1-\omega)m(t)]\delta(p) - [\omega +(1-\omega)m(t)]\delta (p-1). \nonumber
\end{eqnarray*}
Integrating by parts the r.h.s. of Eq.~(\ref{dfdt}) and regrouping terms leads to 
\begin{eqnarray*}
&&\int_0^1 \phi(p) \bigg\{  \frac{1}{h} \frac{\partial}{\partial t} f(p,t) 
+ \frac{\partial}{\partial p} \Big\{ \left[ v(p,t)+u(p,t) \right] f(p,t) \Big\} \bigg\} dp \\ 
&& = \left[ v(1,t)+u(1,t) \right] f(1,t) \phi(1) \\
&& - \left[ v(0,t)+u(0,t) \right] f(0,t) \phi(0). 
\end{eqnarray*}
Since this relation holds for any function $\phi$ we see that $f$ satisfies formally the transport equation 
\begin{eqnarray}
\frac{\partial}{\partial t} f(p,t) &=& - \frac{\partial}{\partial p} \Big\{ h \left[ v(p,t)+u(p,t) \right] f(p,t) \Big\}  \nonumber \\
  &+& h \left[ v(1,t)+u(1,t) \right] f(1,t)\delta(p-1) \nonumber \\
  &-& h \left[ v(0,t)+u(0,t) \right] f(0,t) \delta(p). 
  \label{dfdt}
\end{eqnarray}
Equation~(\ref{dfdt}) expresses the conservation of total number of particles under the transport induced by the effective drift $v+u$ and with source terms
$ h \left[ v(1,t)+u(1,t) \right] f(t,1)\delta(p-1)$ and $- h \left[ v(0,t)+u(0,t) \right] f(t,0)\delta(p)$ at the boundary points $p=1$ and $p=0$, respectively.  An intuitive interpretation of this equation is that the mass density $f(p,t)$ is transported by the field $v$ in $[0,1]$ and suffers and additional impulse at the borders $p=0,1$ given by the field $u$, which is associated to the rate Eqs.~(\ref{dndt0}) and (\ref{dndtS}) for $n_0$ and $n_S$, respectively.  This is reminiscent of the bouncing effect of particles at the boundaries, by which a particle that hits $p=0$ ($p=1$) can later jump back to the interval $(0,1)$ with probability $(1-\omega)m(t)=v(0,t)+u(0,t)$ [$(1-\omega)(1-m)=-v(1,t)-u(1,t)$].

\subsection{Approximate stationary state solution}
\label{approximate}

It is useful to decompose $f(p,t)$ into a sum of a boundary term $f(0,t) \, \delta(p) + f(1,t) \, \delta(p-1)$ taking into account the dynamics near $p=0$ and $p=1$, and an inside term $\tilde f(p,t)$ that describes the dynamics in $(0,1)$:
\begin{eqnarray*}
  f(p,t) = \tilde f(p,t) + f(0,t) \, \delta(p) + f(1,t) \, \delta(p-1),
\end{eqnarray*}
with $\tilde f(0,t) = \tilde f(1,t) = 0$.  Then, Eq.~(\ref{dphidt}) becomes 
\begin{eqnarray}
  \frac{1}{h}\frac{d}{dt} \langle \phi \rangle_f &=& 
  \Big \langle v \, \phi' + \frac{h}{2} \phi'' \Big \rangle_{\tilde f} + B_0 \, f(0,t) \left[ \phi'(0) - \frac{h}{2} \phi''(0) \right] \nonumber \\ &-& B_1 \, f(1,t) \left[ \phi'(1) - \frac{h}{2} \phi''(1) \right],
  \label{dphidt-2}
\end{eqnarray}
where we have neglected terms of order $2$ and higher, and simplified the notation by writing $v=v(p,t)$, $B_0=B_0(t)$, $B_1=B_1(t)$ and $\phi=\phi(p)$.  We are interested in  the stationary solutions to Eq.~(\ref{dphidt-2}).  As expected from previous results, the consensus states $f^*(p)=\delta(p)$ and $f^*(p)=\delta(p-1)$ are stationary solutions.  One can check that by noticing that for $p=0$ ($p=1$) consensus is $\tilde f^*(p)=0$, $f^*(0)=1$ ($f^*(0)=0$), $f^*(1)=0$ ($f^*(1)=1$) and $m=0$ ($m=1$).  In view of our findings in section~\ref{rate} we also expect a symmetric polarized state with mean $m^*=1/2$ to be a stationary solution.  In that perspective it makes  sense to drop the terms involving $\phi''(0)$ and $\phi''(1)$ for symmetry reasons.  We then look for a stationary solution
$\tilde f^*(p) = f^*(p) - f^*(0) \, \delta(p) - f^*(1) \, \delta(p-1)$ satisfying 
\begin{equation}
  \Big \langle v \, \phi' + \frac{h}{2}\phi'' \Big \rangle_{\tilde f^*}  
  + B_0 \, f^*(0) \, \phi'(0)  - B_1 \, f^*(1) \, \phi'(1) =0,
  \label{f-stat}
\end{equation} 
for any $\phi(p)$.  In Appendix~\ref{stationary-f} we show that the solution to Eq.~(\ref{f-stat}), different from $\delta(p)$ and $\delta(p-1)$, is given by 
\begin{eqnarray}
\label{f-stat-1}
  f^*(p) &=& A \bigg\{ \exp \left[\frac{2 \, \omega}{h} \left( p-\frac{\alpha}{2 \, \omega} \right)^2 \right] \\
  &+& \frac{h}{2(1-\omega)} \exp \left( \frac{\alpha^2}{2 \, \omega \, h} \right) \left[ \frac{\delta(p)}{m} + \frac{\delta(p-1)}{1-m} \right] \bigg\},  \nonumber
\end{eqnarray} 
where $\alpha \equiv 1-2(1-\omega)m$ and $A>0$ is a normalization constant that satisfies the condition $\int_0^1 f^*(p) \, dp =1$.  Notice that the magnitude $m$ in Eq.~(\ref{f-stat-1}) is the mean propensity that must satisfy the relation $\int_0^1 (p-m) \, f^*(p) dp = 0$, which is equivalent to 
\begin{eqnarray*} 
  \int_0^1 (p-m) \, \exp \left[\frac{2 \, \omega}{h} \left( p-\frac{\alpha}{2 \, \omega} \right)^2 \right] \, dp = 0. 
\end{eqnarray*} 
This is a nonlinear equation in $m$ that we studied numerically for various values of $\omega$ and $h$.  We found that $m=1/2$ is the only solution in all cases, which is in agreement with the symmetric solution of the rate Eqs.~(\ref{dndt}) in section~\ref{rate}.  Therefore, the symmetric stationary distribution of Eq.~(\ref{dphidt-2}) is given by
\begin{eqnarray}
  \label{f-stat-1}
  f^*(p) &=& A \bigg\{ \exp \left[\frac{2 \, \omega}{h} \left( p-\frac{1}{2} \right)^2 \right] \nonumber \\
  &+& \frac{h}{(1-\omega)} \exp \left( \frac{\omega}{2 \, h} \right) \left[ \delta(p) + \delta(p-1) \right] \bigg\}, 
\end{eqnarray} 
with
\begin{eqnarray*}
A = \Bigg\{ \int_0^1 \exp \left[\frac{2 \, \omega}{h} \left( p-\frac{1}{2} \right)^2 \right] dp 
  + \frac{2h \, \exp \left( \frac{\omega}{2 \, h} \right) }{(1-\omega)} \Bigg\}^{-1}.
\end{eqnarray*}
As we can see, $f^*(p)$ is symmetric around $m=1/2$ and is the sum of the continuous function $A \exp \left[ \frac{2\omega}{h} \left( p-\frac{1}{2} \right)^2 \right]$ in the interval $(0,1)$ that has the shape of an inverted Gaussian, and the two Dirac masses located at the boundaries, what makes $f^*(p)$ a discontinuous function at $p=0$ and $p=1$.

\begin{figure}[t]	
  \includegraphics[width=\columnwidth]{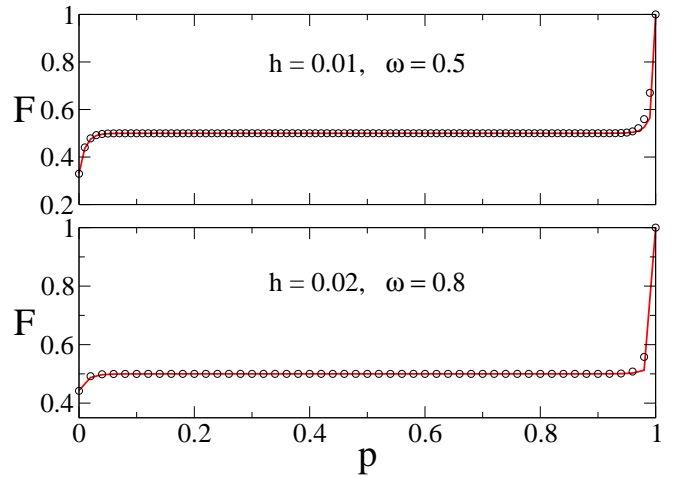}
  \caption{Propensity cumulative distribution vs $p$ for the parameter values indicated in the legends.  Solid lines correspond to the approximate continuous solution for $h \ll 1$ from Eq.~(\ref{F-stat}), while circles represent the exact discrete solution $F_k^*=\sum_{k'=0}^k n_{k'}$.}
  \label{F-p}
\end{figure}

We can alternatively describe the stationary solution by the cumulative distribution function of $f^*(p)$ 
\begin{equation}
  F^*(p)= \begin{cases} 
    0 ~~ \mbox{if  $p<0$}, \\ 
    \frac{A \, h}{(1-\omega)} \exp \left( \frac{\omega}{2h} \right) ~~ \mbox{if  $p=0$}, \\ 
    A \int_0^p e^{ \frac{2\omega}{h} \left( p-\frac{1}{2} \right)^2 } dp + \frac{A \, h}{(1-\omega)} \exp \left( \frac{\omega}{2h} \right) ~~ \mbox{if $0\le p < 1$}, \\ 1 ~~ \mbox{if $p \ge 1$}.
  \end{cases}
  \label{F-stat}
\end{equation}
In Fig.~\ref{F-p} we plot the approximate stationary cumulative distribution $F^*(p)$ for continuous $p$ (solid curves) and the exact discrete cumulative distribution $F_k^*=\sum_{k'=0}^k n_{k'}$ (circles) for two different small values of $h$.  We see that the data for $F^*(p)$ agrees very well with that of $F_k^*$, showing that $f^*(p)$ given by Eq.~(\ref{f-stat-1}) is indeed the limit of $\sum_{k=0}^S n_k^* \, \delta(p-kh)$ when $h \to 0$.

\section{Summary and discussion}
\label{summary}

We studied a system of interacting particles that models the dynamics of voting intentions in a population of individuals that interact by pairs.  The propensity of an individual to vote for a given candidate may either increase or decrease after interacting with other partner, depending on the propensity of the partner and the weight $\omega$ in $[0,1]$ assigned to its own propensity.  We have investigated the dynamics of the system by means of a rate equation approach and we have checked the results with MC simulations.  Starting from a nearly uniform distribution of propensities in $[0,1]$, we found that for $\omega=0$ the system is quickly driven towards an extreme propensity ($p=0$ or $p=1$) that corresponds to the initial majority.  The dynamics stops evolving when all individuals share the same extreme propensity; an absorbing consensus state.  However, for $\omega>0$ the evolution is quite different: the system initially evolves towards a stationary state characterized by a distribution of propensities that is symmetric around $p=1/2$ and peaked at the extreme values $p=0$ and $p=1$, and it becomes more pronounced when $\omega$ gets larger.  This distribution describes a state of polarization where most individuals adopt extreme values of $p$, whose effect is magnified as $\omega$ increases. This implies that a tiny weight assigned to our own propensity is enough to polarize the population into two groups with extreme and opposite propensities.  However, this state of symmetric polarization is unstable, and thus any perturbation from that state leads the system towards one of the two extremist consensus.  Single MC simulations of the dynamics of the model showed that, indeed, the system may initially reach this symmetric quasi-stationary state but finite-size fluctuations eventually drive the system towards one of the two absorbing configurations.   An stability analysis of the rate equations shows that any symmetric distribution evolves towards the polarized state, but there are also non-trivial propensity distributions with mean propensity $m=1/2$ that are not symmetric around $p=1/2$ and that evolve and reach the polarized state.

An insight into the polarized state was obtained by analyzing the continuous limit of the system of rate equations.  This approximation lead to a transport equation with a convection term that represents a drift of particles from the center propensity $p=1/2$ towards the extremes, which induces polarization.  The stationary solution has the shape of an inverted Gaussian with two Delta functions at $p=0$ and $p=1$ that account for the dynamics at the boundaries.  In this peculiar dynamics, particles can hit and stay at one of the boundaries for some time but eventually leave, and then hit the boundary again and so on, following and endless loop.  We have quantified the lifetime of the polarized state by measuring the mean consensus time $\tau$, and found that it increases with $\omega$ and diverges as $\tau \sim (1-\omega)^{-2}$ when $\omega$ approaches $1$.  This would imply that polarization is quite stable in populations with narrow-minded individuals that only take into account its own opinion when interacting with others, reinforcing their previous believes and adopting more extreme viewpoints.  This result is akin to that obtained in related models for opinion formation \cite{Mas-2013,Mas-2013-2,LaRocca-2014,Velasquez-2018} that include a reinforcement mechanism by which pairs of individuals with the same opinion orientation (both in favor or both against a given political issue) are more likely to interact and become more extremists.  Even though the propensity model studied in this article does not include this mechanism implicitly, it is able to capture the same phenomenology by implementing a simple interaction rule
that consider pairwise interactions between individuals as independent of the  opinion group they belong to.  

In the studied model, the propensity update probability is a simple weighted average of the propensities of the two interacting individuals.  It would be worthwhile to explore some extensions of the model that consider updating probabilities that are non-linear functions of the propensities and investigate how the behavior of the model is affected, for instance, whether the polarized state becomes more stable or not.  It might also be interesting to study versions of the model where pairwise interactions are not simply taken as all-to-all, but rather take place on lattices or complex networks.  These are all topics for future investigation.

\onecolumngrid

\appendix

\section{Stationary solutions of the rate equations} 
\label{stat-disc}

Setting the time derivatives of Eqs.~(\ref{dndt}) to zero leads to the system 

\begin{subequations}
  \begin{alignat}{3}
    \left[ 1-\omega h - (1-\omega) m) \right] n_1 -(1-\omega) m \, n_0 & =  0, 
    \label{n-statn0} \\
    \left[ \omega h (k-1)+(1-\omega) m\right] n_{k-1} - n_k 
    +  \big\{ 1 - \left[ \omega h (k+1)+ (1-\omega) m \right] \big\} n_{k+1} 
    & =  0 \qquad 1 \le k \le S-1, \label{n-statnk}  \\
    \left[ \omega (1-h) + (1-\omega) m \right] n_{S-1} - (1-\omega) (1-m) n_S &  =  0.
    \label{n-statnS} 
  \end{alignat}
  \label{n-statEqu}
\end{subequations}
Notice that the two consensus states corresponding to 
\begin{itemize} 
\item $m=0$, $n_0=1$, $n_k=0$ for $k=1,..,S$, 
\item $m=1$, $n_1=1$, $n_k=0$ for $k=0,..,S-1$,
\end{itemize}
are solutions of the system of Eqs.~(\ref{n-statEqu}).  To find other possible non-trivial solutions we first note that $n_k$ (for all $k=1,..,S$) can be expressed as a function of $n_0$.  Starting from Eq.~(\ref{n-statn0}) we obtain
\begin{equation}
n_1 = \frac{(1-\omega) m n_0}{1-\omega h - (1-\omega) m}.
\label{n1-n0}
\end{equation}
Then, solving for $n_2$ from Eq.~(\ref{n-statnk}) for $k=1$, and using the previous  expression for $n_1$ we obtain 
\begin{equation}
n_2 = \frac{mn_0(1-\omega)(\omega h+(1-\omega)m)}
{(1-\omega h-(1-\omega) m)(1-2\omega h-(1-\omega)m)}.
\end{equation}
The same procedure applied to $k=2$ leads to
\begin{equation}
n_3 = \frac{mn_0(1-\omega)(\omega h+(1-\omega)m)(2\omega h+(1-\omega)m)}
{(1-\omega h-(1-\omega) m)(1-2\omega h-(1-\omega)m)(1-3\omega h-(1-\omega)m)}.
\end{equation}
In general  we have  
\begin{equation}\label{n-statk}
n_k = n_0 \, G_k(h,\omega,m) \qquad 1 \le k \le S, 
\end{equation}
where
\begin{eqnarray}
G_k(h,\omega,m) 
& = & 
\frac{\Pi_{j=0}^{k-1} \left[ (1-\omega)m+j\omega h \right]}
{\Pi_{j=1}^k \left[ 1-(1-\omega)m -j\omega h \right]} \label{gkm}, 
\end{eqnarray}
as quoted in Eq.~(\ref{gkmw}) of the main text.  The value of $n_0$ can be obtained by inserting the expression $n_k = n_0 \, G_k(h,\omega,m)$ in the normalization condition $\sum_{k=0}^S n_k=1$ and solving for $n_0$, which leads to the expression
\begin{equation}\label{n-stat0}
n_0 = \frac{1}{1+\sum_{k=1}^S G_k(h,\omega,m)}.
\end{equation}
quoted in Eq.~(\ref{n-stat}) of the main text.
Using relation \ref{n-statk} we can rewrite the mean propensity $m$ as 
\begin{eqnarray*}
	m =  \sum_{k=0}^S hk n_k = h n_0 \sum_{k=1}^S k \, G_k(h,\omega,m),
\end{eqnarray*}
which after replacing the expression \ref{n-stat0} for $n_0$ and rearranging the terms becomes Eq.~(\ref{m-stat}) of the main text. 

When $m>0$ we can rewrite $G_k(h,\omega,m)$ in terms of the gamma functions by using the Pochhammer formula
\begin{equation}
(z)_k \equiv z(z+1)(z+2)...(z+k-1)=\frac{\Gamma(z+k)}{\Gamma(z)} \qquad
\mbox{for $z\in\mathbb{C}\backslash \mathbb{Z}_-$ and $k \ge 0$ integer}, 
\end{equation}
which follows from the relation $\Gamma(z+1)=z \, \Gamma(z)$.  We first rewrite the numerator of $G_k(h,\omega,m)$ in Eq.~(\ref{gkm}) introducing $z \equiv (1-\omega)m/(\omega h)$ as 
\begin{eqnarray*} 
	&& (1-\omega)m(\omega h)^{k-1}\Big(\frac{(1-\omega)m}{\omega h}+1\Big)
	\Big(\frac{(1-\omega)m}{\omega h}+2\Big)...\Big(\frac{(1-\omega)m}{\omega h}+k-1\Big)  
	= (\omega h)^k z(z+1)...(z+k-1) \\ 
	&& = (\omega h)^k \frac{\Gamma(z+k)}{\Gamma(z)}.  
\end{eqnarray*} 
Notice that if $m=0$ then $z=0$ and we cannot use Pochhammer formula.  Letting $\tilde z \equiv \left[1-(1-\omega)m - k\omega h \right]/(\omega h)$, we rewrite in the same way the denominator of 
$G_k(h,\omega,m)$ in Eq.~(\ref{gkm}) as 
\begin{eqnarray*} 
	&& (\omega h)^k \tilde z(\tilde z+1)...(\tilde z+k-1) 
	= (\omega h)^k \frac{\Gamma(\tilde z+k)}{\Gamma(\tilde z)}.  
\end{eqnarray*} 
Inserting these two last expressions for the numerator and denominator of $G_k(h,\omega,m)$ in Eq.~(\ref{gkm}) leads to the expression quoted in Eq.~(\ref{gkmw-G}) of the main text.

\section{Continuum equation for $\langle \phi \rangle_f$} 
\label{continuum-eq}
 
 In this section we derive an equation for the time evolution of the mean of a generic function $\phi(p)$ over the population of agents, expressed as  
 \begin{eqnarray}
  \langle \phi \rangle_f(t) \equiv \int_0^1 \phi(p) f(p,t) dp = \sum_{k=0}^S n_k(t) \, \phi(kh),
	\label{phi-ave}
\end{eqnarray}
where $f(p,t)$ is the propensity distribution at time $t$.  Taking the time derivative of Eq.~(\ref{phi-ave}) gives
\begin{eqnarray} 
\label{dphi-ave-dt}
\frac{d}{dt} \langle \phi \rangle_f  &=& \sum_{k=0}^S \frac{d n_k}{dt}  \phi(kh) \nonumber \\ 
&=& A\phi(0)+B\phi(1) - \sum_{k=1}^{S-1} n_k\phi(kh) 
 +  \sum_{k=1}^{S-1} n_{k-1} \phi(kh) [\omega h (k-1) + (1-\omega)m] \\
 &+& \sum_{k=1}^{S-1} n_{k+1} \phi(kh) [1-\omega h(k+1) - (1-\omega)m], \nonumber 
\end{eqnarray} 
with 
\begin{eqnarray*}
A \equiv (1-\omega h - (1-\omega)m)n_1 - (1-\omega)mn_0 ~~~ \mbox{and} ~~~
B \equiv (\omega (1-h) + (1-\omega)m) n_{S-1} - (1-\omega)(1-m)n_S. 
\end{eqnarray*} 
We then write 
\begin{eqnarray*}
	\phi(kh) &=& \phi((k-1)h) + h\phi'((k-1)h) + \frac{h^2}{2}\phi''((k-1)h) + O(h^3)  ~~~\mbox{and} \\
	\phi(kh) &=& \phi((k+1)h) - h\phi'((k+1)h) + \frac{h^2}{2}\phi''((k+1)h) + O(h^3), 
\end{eqnarray*} 
and replace these expressions in the summations of Eq.~(\ref{dphi-ave-dt}). 
Then 
\begin{equation} \label{dphi-ave-dt-2}
\begin{split} 
 \frac{d}{dt} \langle \phi \rangle_f & = 
(A+n_0)\phi(0)+(B+n_S)\phi(1) - \langle \phi \rangle_f  \\ 
& +  \sum_{k=0}^{S-2} n_k \left[ \omega h k + (1-\omega)m \right] 
                               	\left[ \phi(kh) + h\phi'(kh) + \frac{h^2}{2}\phi''(kh) \right]  \\
 & + \sum_{k=2}^{S} n_k \left[1-\omega hk - (1-\omega)m \right]
 \left[ \phi(kh) - h\phi'(kh)  + \frac{h^2}{2}\phi''(kh) \right] + O(h^3). 
 \end{split} 
\end{equation} 
The first summation in the rhs of Eq.~(\ref{dphi-ave-dt-2}) is 
\begin{eqnarray*} 
&& \sum_{k=0}^S n_k  [\omega h k + (1-\omega)m]  
                    \left[ \phi(kh) + h\phi'(kh) + \frac{h^2}{2}\phi''(kh) \right]  \\
&& - n_{S-1} [\omega (1-h) + (1-\omega)m]  
  \left[ \phi(1-h) + h\phi'(1-h) + \frac{h^2}{2}\phi''(1-h) \right]  \\
&& - n_S [\omega  + (1-\omega)m]  \left[ \phi(1) + h\phi'(1) + \frac{h^2}{2}\phi''(1) \right]  
\end{eqnarray*} 
which is, up to $\mathcal O(h^3)$, 
\begin{equation}\label{EquLimit1}
\begin{split}  
& \Bigg \langle [\omega x + (1-\omega)m] -\left( \phi+h\phi'+\frac{h^2}{2}\phi'' \right) \Bigg \rangle_f
 - \phi(1) \left[ [\omega + (1-\omega)m]n_S + [\omega (1-h)+(1-\omega)m]n_{S-1} \right]  \\ 
&  - h\phi'(1)n_S [\omega + (1-\omega)m] 
- \frac{h^2}{2}\phi''(1)n_S [\omega + (1-\omega)m].
\end{split} 
 \end{equation} 
 The second summation in the rhs of Eq.~(\ref{dphi-ave-dt-2}) is 
\begin{equation*}
\begin{split}  
& \sum_{k=0}^{S} n_k [1-\omega hk - (1-\omega)m] \left[ \phi(kh) - h\phi'(kh) 
 + \frac{h^2}{2}\phi''(kh) \right] \\ 
&  - n_0 [1-(1-\omega)m] \left[ \phi(0) - h\phi'(0)  + \frac{h^2}{2}\phi''(0) \right] \\ 
&  - n_1 [1-\omega h - (1-\omega)m]  \left[ \phi(h) - h\phi'(h)  + \frac{h^2}{2}\phi''(h) \right],
\end{split} 
 \end{equation*} 
which is, up to $\mathcal O(h^3)$, 
\begin{equation}\label{EquLimit2}
\begin{split}  
&  \Bigg \langle (1-\omega x - (1-\omega)m) \left( \phi - h\phi'+\frac{h^2}{2}\phi'' \right) \Bigg \rangle_f
- \phi(0) \left[ n_0[1-(1-\omega)m]  - n_1[1-\omega h - (1-\omega)m] \right] \\ 
& + h\phi'(0)n_0 [1-(1-\omega)m] - \frac{h^2}{2}\phi''(0)n_0[1-(1-\omega)m].
\end{split} 
 \end{equation} 
Replacing the two summations in Eq.~(\ref{dphi-ave-dt-2}) by the expressions (\ref{EquLimit1}) and (\ref{EquLimit2}) leads to Eq.~(\ref{dphidt}) quoted in the main text.

\section{Stationary solution of the equation for $\langle \phi \rangle_f$} 
\label{stationary-f}

We look for a stationary solution $f^*$ of the form 
\begin{eqnarray*}
	f^*(p) = \tilde f^*(p) + f^*(0) \delta(p)+ f^*(1) \delta(p-1),
\end{eqnarray*}	
where $\tilde f^*(p)$ is a continuous function of $p$.  We can then rewrite Eq.~(\ref{f-stat}) as 
\begin{equation*}
	\begin{split} 
	0 & = 
	\int_0^1 (2\omega p+2(1-\omega)m-1) \phi'(p) \, \tilde f^*(p) \,dp  
	+ \frac{h}{2} \int_0^1 \phi''(p) \, \tilde f^*(p) \,dp  \\ 
	&+ \phi'(0) f^*(0)(1-\omega)m - f^*(1) (1-\omega)(1-m) \phi'(1).
	\end{split} 
\end{equation*} 
Integrating by parts the second integral of the above equation gives
\begin{equation*}
\begin{split} 
0 & = 
\int_0^1 \phi'(p) \Big\{ (2\omega p+2(1-\omega)m-1) \tilde f^*(p) - \frac{h}{2} \tilde f^{*'}(p) \Big\} \,dp   \\ 
&+ \phi'(0) \Big\{ f^*(0)(1-\omega)m - \frac{h}{2} \tilde f^*(0)\Big\} 
+ \phi'(1) \Big\{\frac{h}{2} \tilde f^*(1) - f^*(1) (1-\omega)(1-m)\Big\}.
\end{split} 
\end{equation*} 
Since this equality must hold for any function $\phi(p)$ we obtain that
\begin{eqnarray*} 
&& \left[ 2\omega p+2(1-\omega)m-1 \right] \tilde f^*(p) - \frac{h}{2} \tilde f^{*'}(p) = 0, \\ 
&& f^*(0)(1-\omega)m = \frac{h}{2} \tilde f^*(0), \\ 
&& \frac{h}{2} \tilde f^*(1) = f^*(1) (1-\omega)(1-m),
\end{eqnarray*} 
from which we arrive to the expression for $f^*(p)$ quoted in Eq.~(\ref{f-stat-1}). 

\begin{acknowledgments}
This work was partially supported by Universidad de Buenos Aires under grants 20020170100445BA, 20020170200256BA, and by ANPCyT PICT2012 0153 and PICT2014-1771.  
\end{acknowledgments}

\bibliographystyle{apsrev}

\bibliography{references}

\end{document}